\documentclass[prd,aps,floats,preprint,nofootinbib,showpacs]{revtex4-1}
\usepackage{amssymb}
\usepackage{amsmath}
\usepackage{mathrsfs}
\usepackage[dvips]{graphicx}
\usepackage[hyperindex]{hyperref}
\hypersetup{breaklinks=true, hyperfootnotes=true, pagecolor=white, colorlinks=true, 
allcolors=blue
}
\usepackage[all]{hypcap}
\usepackage{mathptmx}

\makeatletter
\DeclareRobustCommand*{\bfseries}{%
  \not@math@alphabet\bfseries\mathbf
  \fontseries\bfdefault\selectfont
  \boldmath
}
\makeatother

\begin{document}
%\preprint{APS/123-QED}

\title{Connecting the Cabbibo-Kobayashi-Maskawa matrix to quark masses}

\

%\begin{center}
\author {M. Novello}\email[Electronic address:]{novello@cbpf.br}
\author{V. Antunes}\email[Electronic address:]{antunes@cbpf.br}
 \affiliation{
Centro de Estudos Avan\c{c}ados de Cosmologia (CEAC/CBPF) \\
Rua Dr. Xavier Sigaud 150, Urca 22290-180 Rio de Janeiro, RJ-Brazil}
\date{\today}% It is always \today, today,


\begin{abstract}
 We show that the Cabbibo-Kobayashi-Maskawa interaction matrix may be constructed with the quark masses.
\end{abstract}

\maketitle

\section{Introduction}

In the standard model (SM) the coupling between quarks and vectorial bosons is provided by the Lagrangian
\begin{equation}
L_{int} = - \frac{g}{\sqrt{2}} ( \bar{u}_{L},  \bar{c}_{L},  \bar{t}_{L} )  \Gamma \, V_{CKM}  \left(
\begin{array}{c}
d_{L} \\
s_{L} \\
b_{L}
\end{array}
\right) + h.c.
\label{ckmint}
\end{equation}
where $\Gamma \equiv \gamma^{\mu} W_{\mu}^{+}$, and the quark mixing is given by the dimensionless Cabbibo-Kobayashi-Maskawa (CKM) matrix \cite{kobayashi_maskawa, cabibbo, pdg_ckm}
\begin{equation}
V_{CKM} =
\begin{pmatrix}
V_{ud} & V_{us} & V_{ub}\\[2pt]
V_{dc}  & V_{sc} & V_{cb}\\[2pt]
V_{dt}  & V_{st}  & V_{bt}
\end{pmatrix}.
\end{equation}
%In the SM the fermions acquire mass through the Yukawa interaction between the Higgs field and the fermions. The role of the Yukawa interaction is indispensable for the Higgs mechanism to provide mass to the quarks.
This description provided by the CKM matrix and the Lagrangian (\ref{ckmint}) has proven very successful in accounting for quark decaying processes, and there are reasonably accurate experimental constraints on the values of the CKM matrix elements, the moduli of which, according to \cite{pdg_ckm}, being approximately
\begin{equation}
|V_{CKM}| \approx
\begin{pmatrix}
0.97 & 0.22 & 3.5 \times 10^{- 3}\\[3pt]
0.22 & 0.97 & 4\times 10^{- 2}\\[3pt]
8.7 \times 10^{- 3} & 4 \times 10^{- 2} & 1
\end{pmatrix}.
 \label{ckm_matrix}
\end{equation}
However, the origin of the CKM matrix remains a mystery in the SM.

In a recent series of papers \cite{novelloCQG,novellobittencourt} one of us has shown how to relate the origin of the masses of all bodies to the Mach principle. We understand this principle, according to ideas explored by Einstein in his theory of gravity, as the statement according to which the inertial properties of a body $B$ are determined
by the energy-momentum throughout all space represented by the most homogeneous state, which is related  to
%the cosmological constant or, in modern language,
the vacuum of all remaining bodies. This amounts to
describe the energy-momentum distribution of all
bodies complementary to $B$ as related to the vacuum energy density
\begin{equation}
T_{\mu\nu} = \rho_{vac}
\, g_{\mu\nu}.
\label{energy_distrib}
\end{equation}
It should be stressed that in this mechanism the gravitational field acts merely as a catalyst, and the final expression of the masses depends neither on the intensity nor on the properties of the gravitational field. Moreover it can be used in a equivalent way to give mass not only to fermions but to all particles, including the Higgs boson \cite{novellobittencourt}.

 In particular, for fermions the gravitational mechanism of mass generation implies the following relation between the mass and the vacuum energy density
 \begin{equation}
 \frac{M}{\sigma} = \frac{\rho_{vac}}{c^{2}},
 \label{30julho13}
 \end{equation}
where $\sigma$ is a constant with dimensions of (length)$^{3}$. According to this expression, the fermion acquires a
mass $M$ that depends crucially on a non-vanishing $\rho_{vac}$. Since $\sigma$ (one for each fermion) has dimensions of (length)$^{3}$, one can define from it an associated effective fermion radius by setting $\sigma = R^3$.
%This is precisely the origin of relation (\ref{hypothesis}) stated above.

A similar situation -- although interpreted in a very distinct way -- occurs in the framework of the Higgs mechanism, where the vacuum state of the Higgs boson has precisely the form of energy distribution as in (\ref{energy_distrib}). In this scenario one can alternatively associate what we call, according to Mach, the rest-of-the-universe to the vacuum of the Higgs field.

In the present letter we will propose an explanation for the origin of the CKM matrix which rests on the gravitational mechanism of mass generation [4, 5]. Only the quark masses are employed here to build the quark mixing matrix, as will be explained in what follows.

%\vspace{0.50cm}
\section{Quark masses and the CKM matrix}

The existence of an effective radius $R_i$ for each fermion $\psi_i$ allows us to make the hypothesis
%(the consequences of which must be confronted with observations) that the interaction matrix $V_{CKM}$ should depend on these radii.  We are then led to
%associate to the modulus of
that
each element $V_{ij}$ of the interaction matrix $V_{CKM}$ is related to the ratios $R_{i}/R_{j}$. Consider the matrix
 \begin{equation}
Q =
\begin{pmatrix}
Q_{ud} & Q_{us} & Q_{ub}\\[2pt]
Q_{dc}  & Q_{sc}  & Q_{cb}\\[2pt]
Q_{dt}  & Q_{st}  & Q_{bt}
\end{pmatrix},
\label{pre_vq}
\end{equation}
 the elements of which are constructed as
\begin{equation}
Q_{ij} \equiv \frac{R_i}{R_j}
= \left( \frac{M_{i}}{M_{j}} \right)^{1/3}, \ \ M_i<M_j ,
\label{ratio}
\end{equation}
where the equality follows from relation (\ref{30julho13}). Let us define the associated matrix
 \begin{equation}
S \equiv
\begin{pmatrix}
 \ Q_{ud} &  \frac{1}{2\Delta}Q_{us} & \ Q_{ub} \ \\[3pt]
 \frac{1}{2\Delta}Q_{dc}   & Q_{sc}  & \ \ Q_{cb}\ \\[3pt]
Q_{dt}  & Q_{st}  & \ \ Q_{bt}\
\end{pmatrix},
\label{pre_vq2}
\end{equation}
where
\begin{equation}
\Delta \equiv \det (Q).
\label{det_q}
\end{equation}
We propose to reformulate the Lagrangian (\ref{ckmint}) by inserting in the place of $V_{CKM}$ the following quark mixing matrix
\begin{equation}
V_{Q} = \exp\big[-i\Delta \big( S + S^{T} \big) \big] ,
\label{expression_e}
\end{equation}
where
\begin{equation}
\exp(A) = \sum_{n=0}^{\infty} \frac{A^n}{n!} \approx I + A
\end{equation}
is the matrix exponential of a matrix $A$, $I$ being the $3\times 3$ identity matrix.
Note that the matrix $V_Q$ is unitary by construction. Note also that no free parameters other then the quark masses are employed here. Interestingly, relation (\ref{expression_e}) bares some resemblance with the matrix exponential parametrization of $V_{CKM}$ proposed in the past \citep{dattoli_sabia_torre, zhukovsky_dattoli}.

According to (\ref{ratio}), a large difference between two quarks masses implies a large ratio between the effective fermion radii associated to them. We have associated the transition rate between two quarks, through charged weak current, to the effective radii overlapping. In this way we can roughly associate to the modulus of each CKM matrix element the ratio between the cubic root of the masses of the two quarks involved in a particular transition.

%Taking into account the current best estimates for quark masses, expression (\ref{expression_e}) leads to a matrix $|V_Q|$ surprisingly similar to $|V_{CKM}|$.
%Indeed,
Adopting here the numerical values for the quark masses as in \cite{pgg_quark_mass}, \textit{viz.}
$M_{t} = 173,1$ Gev,  $M_{b} = 4.18$ Gev,  $M_{c} = 1.28$ Gev,  $M_{s} =  96$ Mev,  $M_{d} = 4.7$ Mev, $ M_{u} = 2.2$ Mev, the determinant of $Q$ is $\Delta = 4.4\times 10^{-2}$, and relations (\ref{expression_e}), (\ref{ratio}) and (\ref{det_q}) yield the moduli matrix
%\begin{widetext}
%\begin{equation}
%V_{Q} \approx
%\begin{pmatrix}
%\ 1 - 7\times 10^{-2}i\ &\ -0.22i\ &\ -5 \times 10^{-3}i\ \\[3pt]
%\ -0.22i\  & \ 1 - 3\times 10^{-2}i\  & \ -3.4\times 10^{-2}i\ \\[3pt]
%\ -5 \times 10^{-3}i\  & \ -3.4\times 10^{-2}i\  & \ 1 - 2.6\times 10^{-2}i\
%\end{pmatrix}.
%\label{vqc}
%\end{equation}
%\end{widetext}
\begin{equation}
|V_{Q}| \approx
\begin{pmatrix}
1 & 0.22 & 5 \times 10^{-3} \\[3pt]
0.22 & 1  & 3.4\times 10^{-2}\\[3pt]
5 \times 10^{-3}  & 3.4\times 10^{-2}  & 1
\end{pmatrix},
\label{vq}
\end{equation}
which has a striking similarity with the CKM matrix (\ref{ckm_matrix}).
The discrepancies we encounter here may be credited to inaccuracies in the quark masses estimates, since
they are strongly dependent on the theoretical model adopted to obtain them. The exception for this argument are the terms $V_{us} = V_{dc}$ which must be divided by $2\Delta$ to match the observed values. The reason behind this is still unknown to us.

\section{Final comments}

    The similarity between the CKM matrix (\ref{ckm_matrix}) and the matrix (\ref{vq}) is remarkable and points in the direction that the interaction process is indeed controlled by the ratio of the effective radius. This shows the existence of an intimate connection between the CKM matrix and the quarks masses. The relationship between quark mass and CKM matrix elements is a long term discussed in literature. In fact  an old speculation by Weinberg \cite{weinberg77} related the Cabibbo angle with the square root of the d and s quarks mass ratio. In this paper we related the CKM matrix elements directly with the effective ratio, and indirectly with, not a square root, but cubic root of the ratio of quarks masses.

It is interesting to note that when the vacuum energy density in equation (\ref{energy_distrib}) is associated with the Higgs vacuum, the right-hand-side of relation (\ref{30julho13}) can be written in terms of the density of the Higgs vacuum energy as
 \begin{equation}
\frac{M_{q} \, c^{2}}{ R^{3}_{q}} = \frac{E^{4}_{vac.}}{(\hbar \, c)^{3}}.
\label{hypothesis2}
\end{equation}
As a consequence, we can obtain the value of the effective fermion radius $ R_{q}$ in terms of its corresponding Compton radius $\lambda_{q}=h/M_qc$ as, \textit{viz.}
\begin{equation}
\frac{R_q}{\lambda_q}  = \frac{1}{2\pi}\left( \frac{E_{q}}{E_{vac.}}\right)^{4/3}.
\end{equation}
It turns out that the quark effective radius is always smaller than its corresponding Compton wavelenght. It should be stressed that for the present analysis relating the CKM matrix to the gravitational mechanism of mass generation it is not necessary to know the value of the vacuum energy $E_{vac.}$.

A  procedure similar to the one presented here for quarks can be employed to the construction of the Pontecorvo-Maki-Nakagawa-Sakata matrix which describes neutrino mixing in the leptonic sector. We will analyze this case elsewhere.
\vspace{0.40cm}
\subsection*{Acknowledgements}
We would like to thank I. Bediaga for valuable insights, comments, and intense participation in the discussions made along the elaboration of this work. 
We also thank the financial support from brazilian agencies FINEP, CAPES, CNPq, and FAPERJ.

\end{document}